\documentclass[12pt,a4paper,titlepage]{article}  
\usepackage{graphicx}
\usepackage{amsmath,amssymb}
\usepackage{color}
\usepackage{ulem}
\begin{document}
\baselineskip=24truept

\title{Electromagnetically induced transparency in systems with degenerate autoionizing levels in $\Lambda$-configuration}

\author{Thuan Bui Dinh\thanks{Quantum Optics and Engineering Division, Institute of Physics, University of Zielona G\'ora, ul. prof. A. Szafrana 4a, 65-516 Zielona G\'ora, Poland}, 
Wies\l aw Leo\'nski$^\dagger$\footnotemark[1],
Van Cao Long\footnotemark[1]
and 
Jan Pe\v{r}ina Jr.\thanks{Institute of Physics of AS CR, Joint Laboratory of
Optics, 17. listopadu 50a, 772 07 Olomouc, Czech Republic \newline\indent$^\dagger$ corresponding author: e-mail \tt{wleonski@proton.if.uz.zgora.pl}}
}
%\thanks{corresponding author}

\maketitle
\begin{abstract}
We discuss a $\Lambda$-like model of atomic levels involving two autoionizing (AI) states of the same energy. The system is irradiated by two external electromagnetic fields (strong -- driving and weak -- probing ones). For such a system containing degenerate AI levels we derive the analytical formula describing the medium susceptibility. We show that the presence of the second AI level lead to the additional electromagnetically induced transparency (EIT) window appearance. We show that the characteristic of this window can be manipulated by changes of the parameters describing the interactions of AI levels with other ones. This is a new mechanism which leads to additional transparency windows in EIT model,  that differs from  the mechanism, where a bigger number of Zeeman sublevels is taken into account.
\end{abstract} 
\maketitle

%\newpage

\section{Introduction}
Electromagnetically induced transparency (EIT)discovered for the first time by Harris and co-workers \cite{IH89,HFI90,BIH91} relies on the destructive quantum interference of the transition amplitudes. Such interference leads to suppression of absorption or even to complete transmission of the resonant weak probe beam. This phenomenon arises in the presence of a second (strong) laser beam coupling coherently one of the states which participate in absorption, with some other atomic state. Some reviews concerning EIT are given in literature (see for instance \cite{FIM05,KCG10} \textit{and the references quoted therein}). EIT, in its classical model, can be observed for three basic atomic levels configurations. They are $\Lambda$-, $V$-type and cascade (ladder) ones. In these basic schemes, a single peak of enhanced transmission, or one transparency window appears. Nevertheless, one can find in the literature schemes in which additional transparency windows can appear. Such models can be potentially applied for slowing down of light pulses at various frequencies \cite{WANG04}. The models allowing for multiple transparency windows generation were proposed and discussed, for example in \cite{KCG09} (for the cascade system) and in \cite{WANG06,PGK11} ($\Lambda$-model) (\textit{and the references quoted therein}). 

EIT phenomena can be discussed not only for the models involving discrete levels but also for those containing continuum ones. In particular, as it was shown in \cite{EZL94,PKK99,RRZ06}, it is possible to create transparency windows for systems with autoioniznig (AI) states, or equivalently, with Fano structured continua. Quite recently, in \cite{DCL12} the model with a single AI level was discussed in this context, and the strictly deterministic control laser field was replaced by so-called \textit{white noise} signal.

AI systems involving discrete levels located above a continuum threshold (AI levels) where considered for the first time in the classical paper by Fano \cite{F61}. Fano diagonalization, based on the Coulomb mixing of AI states with the continuum, leads to a nontrivial structure of the latter \cite{RE81,JRC93,DPG01} (\textit{and the references quoted therein}). Such structure can be even more complicated, leading to non-trivial effects in the photoelectron spectra, if we assume that the AI system interacts with other ones, not necessarily containing AI states \cite{PLL11,PLP11}. Such models can lead to the quantum entanglement generation, as well \cite{LPL12}. It should be stressed out that the models involving structured continuum (or continua) described by the Fano profiles play an essential role in various physical processes, and have also a considerable practical meaning. Since first discussions concerning Fano models in atomic physics \cite{F61}, the Fano profile has been found in several functioned materials as plasmonic nanoparticles, quantum dots, photonic crystals and electromagnetic metamaterials -- for exemplary considerations concerning these problems see \cite{TB07,RSF10} \textit{and the references quoted therein}. Discussions concerning those special properties associated with its asymmetric lineshape give us potential applications in a wide range of technologies \cite{Luk10}. An interesting review on Fano profiles in nanostructures in given in \cite{MFK10}.

In this paper, we present  a model comprising continuum states in which interference between two autoionization channels lead to the appearance of additional transparency EIT windows. The  mechanism presented here differs from that for the systems involving only discrete levels without continua. In particular, we shall show that the presence of additional AI states can lead to new quantum interference effects. As a result, the additional EIT windows appear. Moreover, changing the parameters corresponding to the transition to (from) AI states, one can manipulate the characteristic of these windows and the distances between them. However, one should keep in mind that for some particular experimental realizations of our model, some difficulties could appear  during adjustment of the parameters involved in the problem.

The model discussed here is an extension of that involving AI resonances  considered by Raczy\'nski \textit{et. al} \cite{RRZ06} by inclusing the second AI level into our considerations. As it was shown in \cite{LTK87,LT88,LB90,L93} the presence of additional AI states can lead to new quantum interference phenomena present in the system. As a result, additional zeros can appear in long-time photoelectron spectra. In this paper, for simplicity, we shall restrict ourselves to the case when two AI levels are of the same energy, \textit{i.e.} are degenerate. We shall show that such interference related to the presence of additional AI level can lead to the additional EIT window appearance. Moreover, changing the parameters corresponding to the transition to (from) AI states, we can manipulate the characteristic of that window. Our model is easily expendable just by adding more than one extra AI state. For such a situation new, supplementary transparency windows will appear.

\section{The model and solution}
In this Section, we extend the $\Lambda$-like model discussed by Raczy\'nski \textit{et.al.} \cite{RRZ06} which contains a single autoionizing level and a flat continuum coupled to other two lower discrete ones by an external laser field. In our model, instead of the one AI level we discuss two AI levels $|a_1\rangle$, $|a_2\rangle$ with the same energy $E_1=E_2$. Moreover, they are embedded in a same flat continuum $|E\rangle$. All of these states are coupled by a weak probe field of frequency $\omega_p$ with a discrete level $|b\rangle$ and by a relatively strong driving control field with frequency $\omega_c$ with another level $|c\rangle$. The scheme of the model is shown in Fig.1.  The configurational coupling between the AI levels $|a_1\rangle$, $|a_2\rangle$ and flat continuum $|E\rangle$ is described correspondingly by the parameters $U_1$ and $U_2$. We call such scheme as a \textit{double}-$\Lambda$ system. 

In the scheme presented here, the coupling between the excited levels $|a_i\rangle$ ($i=\{1,2\}$) and  the lower discrete ones ($|b\rangle$ and $|c\rangle$) is implemented by external laser fields. In particular, the state $|b\rangle$ is coupled to the continuum $|E\rangle$ and AI levels by a weak probe field with amplitude $\varepsilon_1$, whereas the state $|c\rangle$ by a control field of amplitude $\varepsilon_2$. It is well-known that non-resonant interactions with other levels leads to level shift. Therefore, the field frequencies (especially, the frequency of the strong driving field $\omega_2$) should be properly chosen for omitting such shift.

We start from the full Hamiltonian for the system
\begin{equation}
\hat{H'}=\hat{H'_{0}}+\hat{H'_{1}}
\end{equation}
where
\begin{subequations}\label{oH}
\begin{align}
\hat{H'_{0}}=&\sum\limits_{k=1,2} \hbar\omega_{a_k}|a_k\rangle\langle a_k|+
\hbar\omega_{b}|b\rangle\langle b| +\hbar\omega_{c}|c\rangle\langle c| +\int dE E|E\rangle\langle E|\\
\hat{H'_{1}}=&\big\{\varepsilon_{1}\int dE\langle E|d|b\rangle e^{i\omega_{p}t}|E\rangle\langle b|+\varepsilon_{2}\int dE\langle E|d|c\rangle e^{i\omega_{c}t}|E\rangle\langle c| \\ \nonumber
&+ \int dE \langle E|U_1|a_1\rangle + \int dE \langle E|U_2|a_2\rangle \big\}+ H.c.,
\end{align}
\end{subequations}
where $d$ corresponds to the electric dipole moment, whereas $\varepsilon_1$ and $\varepsilon_2$ describe strengths of the probe and control fields, respectively. Moreover,  the matrix elements $\langle E|U_i|a_i\rangle$ ($i=1,2$) describe configuration interaction between AI levels and flat continuum states.  
As usually in the papers concerning AI, we suppose now that the energies of the AI levels and the laser field frequencies are considerably higher than the energy of the threshold of the continuum. In consequence,  we can neglect all threshold effects and in consequence, all integrals over the energies, appearing here will be extended over the entire real axis. Moreover, we assume that all matrix elements appearing in (\ref{oH}) and corresponding to the transitions to (from) the flat continuum are energy independent. 

It is possible to replace the subset of flat continuum states $\{|E\rangle\}$ and coupled to them AI levels $|a_i\rangle$ ($i=1,2$) by continuum states $|E)$ (denoted here by symbols with round brace , contrary to the flat ones labeled by the usually used symbols $|E\rangle$) with some structure (density function). This function can be derived with use of \textit{Fano diagonalization} method proposed in \cite{F61} and then developed, for instance in \cite{DPG01}). This procedure leads to the scheme in which discrete levels $|b\rangle$ and $|c\rangle$ are coupled to the excited continuum characterized by some density of states. This density function is referred as to \textit{double Fano profile}. Its shape is determined by the ratio between the matrix elements corresponding to the transitions from (to) a some considered discrete level $|j\rangle$ to (from) a flat and structured continua~\cite{F61,LTK87}:
\begin{equation}\label{eq1}
\frac{\langle j|d| E)}{\langle j|d|E\rangle}\,=\, \frac{(E-E_{1})(E-E_{2})+E(q_{1j}\gamma _{1}+ q_{2j}\gamma _{2})-(E_{1} q_{2j} \gamma _{2} +E_{2} q_{1j} \gamma _{1})}{(E-E_{1})(E-E_{2})-iE(\gamma _{1} +\gamma _{2})+i(E_{1} \gamma _{2} +E_{2} \gamma _{1})}\quad ,
\end{equation}
where the widths $\gamma_{1}=\pi|\langle a_1 |U_1|E\rangle |^2$ and $\gamma_{2}=\pi|\langle a_2 |U_2|E\rangle |^2$  are autoionization widths of AI states. Similarly as in \cite{LTK87}, we defined \textit{Fano asymmetry parameters} $q_{1j}$ and $q_{2j}$ which can be expressed as:
\begin{subequations}\label{eq2}
\begin{align}
q_{1j} &=\frac{\langle j|d|a_{1}\rangle }{\pi \langle j|d|E\rangle \langle E|U|a_{1}\rangle}\quad ,\\ % P\int {\frac{\langle j|d| E'\rangle \langle E'| U_{1}| a_{1}\rangle}{E-E'}dE'}
q_{2j} &=\frac{\langle j|d|a_{2}\rangle }{\pi \langle j|d|E\rangle \langle E|U| a_{2}\rangle} \quad ,  % P\int {\frac{\langle j|d| E'\rangle \langle E'| U_{2}| a_{1}\rangle}{E-E'}dE'}
\end{align}
\end{subequations} 
where $j = b, c$. 
These parameters describe ratios between the direct transition between one of the lower states and AI state and its counterpart via (flat) continuum state. It follows from the form of (\ref{eq2}) that when the direct ionization is negligible, the values of the $q$-parameters become high.
 
After performing the diagonalization procedure, the system can be described by following Hamiltonian:
\begin{equation}
\hat{H''}=\hat{H''_{0}}+\hat{H''_{1}}
\end{equation}
where
\begin{subequations}\label{nH}
\begin{align}
\hat{H''_{0}}=&\hbar\omega_{b}|b\rangle\langle b| +\hbar\omega_{c}|c\rangle\langle c| +\int dE E|E)( E|\\
\hat{H''_{1}}=&\big\{\varepsilon_{1}\int dE(E|d|b\rangle e^{i\omega_{p}t}|E)\langle b|+\varepsilon_{2}\int dE(E|d|c\rangle e^{i\omega_{c}t}|E)\langle c|\big\} + Hc.
\end{align}
\end{subequations}
In this formula, all excited levels considered here are replaced by structured continuum states $|E)$.

For further study, we derive the appropriate equations for the density matrix $\rho$. For this purpose, we use Liouville-von Neumann equation and apply the \textit{rotating wave approximation} (RWA) \cite{AE75} which allows to remove rapidly rotating terms from our set of equations. This procedure leads to the following differential equations for the matrix elements of the density matrix $\rho$:
\begin{subequations}\label{eq3}
\begin{align}
i\,\hbar\dot{\rho}_{Eb}&=\left(E-E_{b} -\hbar \omega _{p} \right)\rho _{Eb}-\frac{1}{2}\varepsilon _{1}\left(E|d|b\right\rangle-\frac{1}{2} \varepsilon _{2} \left(E|d|c\right\rangle
 \rho _{cb} \\
i\,\hbar\dot{\rho}_{cb}&=\left(E + \hbar\omega_{c} - E_{b} - \hbar\omega_{p} -i\hbar\gamma_{cb}\right)\rho_{cb}-\frac{1}{2}\varepsilon_{2}^*\!\int{\!\!\langle c | d | E )\rho_{Eb} dE}  \quad ,
\end{align}
\end{subequations}
where the Fano diagonalization formalism was applied.
The above equations are valid within the first order perturbation with respect to the probe field $\varepsilon_1$. The parameter $d$ appearing here is the electric atomic dipole moment and the matrix elements  $\rho_{Eb}=(E|\rho|b\rangle$ and $\rho_{Ec}=(E|\rho|c\rangle$. Moreover, similarly as in \cite{RRZ06}, we have introduced the width  $\gamma_{cb}$. It is a phenomenological relaxation rate for the coherence $\rho_{cb}$.

In general, it is possible to find the full solution of the differential equations (\ref{eq3}), but we shall restrict our considerations to the long-time limit and find the steady-state solution following the way described in \cite{RRZ06}. To solve eqns. (\ref{eq3}), first we eliminate $\rho_{cb}$ expressing it in terms of $\rho_{Eb}$ to get the  integral equation which will be solved in the next step, and next, $\rho_{Eb}$ will be found.

Since, we are interested in EIT, we should calculate the component of the electric polarization of the irradiated medium. It can be expressed as a function of the matrix element $\rho_{Eb}$ in the following way
\begin{eqnarray}\label{eq4}
P^{+}(\omega _{p})\!&=&N\!\int \langle b|d|E)\,\rho_{Eb} \,dE \nonumber\\
 &=&\!-{N}\varepsilon_{1} \left(
 R_{bb} +\frac{\frac{1}{4}\varepsilon_{2}^{2} R_{bc} R_{cb}}{E_{b} +\hbar \omega_{p} -E_{c} -\hbar \omega_{c} -i\hbar \gamma_{cb} -\frac{1}{4} \varepsilon_{2}^{2} R_{cc} } 
 \right)\,=\,\epsilon_{0}\, \varepsilon_{1}\,\chi(\omega_{p}) \quad
\end{eqnarray}
where $N$ is the atom density, $\epsilon_{0}$ is the vacuum electric permittivity, whereas $\chi$ is the medium susceptibility. In our model, the last 
can be expressed as \cite{RRZ06}
\begin{equation}\label{eq5}
\chi(\omega_{p}) =-\frac{N}{\epsilon_{0} } \left(R_{bb} +\frac{\frac{1}{4}\varepsilon_{2}^{2} R_{bc} R_{cb} }{E_{b} +\hbar \omega_{p} -E_{c} -\hbar \omega_{c} -i\hbar \gamma_{cb} -\frac{1}{4} \varepsilon_{2}^{2} R_{cc}}\right)
\end{equation}
with
\begin{equation} \label{eq6}
R_{jk}(\omega _{p})\,=\,\lim_{\substack{\eta\rightarrow 0^{+}\\ \Delta E \rightarrow 0^{+}}}\int \frac{\langle j|d|E)(E|d|k\rangle}{E_{b} -E+\hbar\omega _{p}+ i\eta}\, dE\quad ,\quad(j,k\,=\,b,c) \quad .
\end{equation}
The limit $\eta\rightarrow 0^+$ assures that the imaginary part of $\chi$ will be greater then zero, whereas $\Delta E=E_{2}-E_1$ tends to zero for the degenerate AI levels.
It is worth noting that the function inside the integral contains matrix elements corresponding to the transitions to (from) the structured continuum $|E)$. Since such elements are energy dependent, we apply the formula (\ref{eq1}) to get the explicit dependence of the integrand on the energy. Thus, we can write
\begin{equation}\label{eq7}
R_{jk}(\omega _{p})\,=\,\lim_{\substack{\eta\rightarrow 0^{+}\\ \Delta E \rightarrow 0^{+}}}D_jD_k\int\,\frac{F_j(E)\,F_k(E)}
{E_{b} -E+\hbar\omega _{1}+ i\eta}\,dE \quad ,\quad j,k\,=\,b,c\quad ,
\end{equation}
where the functions $F_j(E),F_k(E)$ inside the integral correspond to matrix elements related to the transitions to (from) the structured continuum $|E)$.
The matrix elements of the dipole moment transition $\langle j|d|E\rangle$ and $\langle E|d|k\rangle$ are denoted by $D_j$ and $D_k$, respectively.
As it was emphasized earlier, we neglect threshold effects, so we extend the integration limits for $R_{jk}(\omega_p)$ from minus to plus infinity. Thanks to this assumption (and other mentioned earlier),  we can find the analytical solution for this parameter and hence, for the medium susceptibility $\chi$.

\section{Results and discussion}
Since we deal with the degenerate case the analytical solution for $R_{jk}(\omega_p)$ can be written as:
\begin{eqnarray}\label{eq9}
R_{jk}=D_{j}D_{k}(Q_{j}+i)(Q_{k}-i)\pi\bigg\{\frac{-i}{(Q_{j}+i)(Q_{k}-i)}-\frac{2i\Gamma}{(Q_{j}+i)(\omega+i\Gamma)}-\frac{2i\Gamma^{2}}{\omega^2+\Gamma^2}\nonumber \\
+\frac{\Gamma}{\omega-i\Gamma}+\frac{\Gamma[{Q_{j21}Q_{k21}+\Gamma_{21}^{2}Q_{j}Q_{k}+\Gamma_{21}(Q_{j21}Q_{k}+Q_{k21}Q_{j}]}}{(1-\Gamma_{21}^{2})(Q_{j}Q_{k}-i(Q_{j}-Q_{k})+1)\omega} \bigg\}
\end{eqnarray}
where the argument was redefined as $\omega\,=\,\hbar\omega_p+E_b-E_1$. Similarly as in \cite{LTK87}, we introduced here the effective asymmetry parameters $Q_{j}, Q_{j21} , \Gamma_{21}$ and AI width $\Gamma$ defined as
\begin{subequations}\label{eq10}
\begin{align}
&Q_{j}=\,\frac{q_{1j}\gamma_{1}+q_{2j}\gamma_{2}}{\Gamma}\quad ,\quad j=c,b\\
&\Gamma=\,\gamma_1\,+\,\gamma_2\quad .
\end{align}
Moreover, we defined the quantiites
\begin{align}
&Q_{j21}\,=\frac{q_{2j}\gamma_{2}-q_{1j}\gamma_{1}}{\Gamma}\quad ,\quad j=c,b\\
&\Gamma_{21}=\frac{\gamma_2-\gamma_1}{\Gamma}.
\end{align}
\end{subequations}

If we assume that the both AI levels are characterized by the same values of parameters describing interaction betwen them and other levels, \textit{i.e.} asymmetry parameters and AI widths, the quantities $\Gamma_{21}=0, Q_{b21}= Q_{c21} =0$. In consequence, our result becomes identical to that derived by Raczynski \textit{et. al} \cite{RRZ06}.

Further, for easier comparison of our results to those presented in \cite{RRZ06} we take the same values for the parameters describing our system as those presented there. Thus, we assume that $\Gamma =10^{-9}$ a.u.,  $D_b=2$ a.u., $D_c=3$ a.u. and the atomic density $N=0.33\times 10^{12}$ cm$^{-3}$. Moreover, the values of the asymmetry parameters are assumed to be  $\sim 10\div100$, whereas the field amplitude $\varepsilon_2$ is within the range from $10^{-9}$ to $10^{-6}$ a.u.  

Thus, Fig. 2 shows the real (dispersion) and imaginary (absorption) parts of the medium susceptibility as a function of the frequency $\omega=\omega_p+(E_b-E_1)/\hbar$ expressed in the units of $\Gamma$. Actually, we see that for the case when $\Gamma_{21}=0, Q_{b21}= Q_{c21}=0$ (solid line), we get the same result as that for the model with a single AI level discussed by Raczy\'nski \textit{et.al} \cite{RRZ06}.  In fact, this is the situation mentioned above, when two degenerate AI levels can be treated as a single  one described by the effective asymmetry parameter and AI width. However, if we assume that the parameters describing two AI levels differ each other, \textit{i.e.} $ Q_{b21}\ne 0$ and  $Q_{c21}\ne 0$, situation changes considerably. For such an additional zero appears in $Im \chi$, leading to the second absorption window occurrence (dashed-dotted line). In consequence, two windows are apparent and they are placed symmetrically with respect to the point $\omega=0$. The position of these windows depends on the values of the parameters describing our system and for some cases the windows coalesce to a single one with a sharp peak inside (dashed line). Such behavior resembles that discussed in \cite{LTK87}, concerning long-time photoelectron spectrum. The second window for our model corresponds to the additional zero in the spectrum discussed in \cite{LTK87} as a result of existence of extra ionization channel via the second AI level. In consequence, an additional quantum interference effect becomes present in the system leading to generation of the second zero in photoelectron spectra and the transparency window, as well. Moreover, for the system considered here one can observe an additional region of anomalous dispersion related to the presence of the second transparency window (as we compare our result with that  discussed in \cite{RRZ06}). Thus, the presence of the second AI state in the system can lead to nontrivial results, analogously to the situation presented in the discussion concerning photoelectron spectra \cite{LTK87,LT88,LB90,L93}. 

The structure of created windows is very clear.  It can be manipulated in potential applications, for example in simultaneous slowing down of light pulses at various frequencies \cite{WANG04}. In particular, the position and widths of the transparency windows can be changed by the strength of control field. In Fig. 3 we show the real and imaginary parts of the medium susceptibility again for various strengths of this field intensity $\epsilon_{2}$. One can see that its changes can influence the positions and widths of the windows. If we increase the value of  $\epsilon_{2}$, both the distance between the windows and their widths increases considerably. Thus, we can use the intensity of the control field as a control parameter for EIT effects.

In Figures 4 and 5, we show how the values of the autoionization widths can influence the number, position and width of the transparency windows. Thus, Fig. 4 corresponds to the situation when the effective asymmetry parameters $Q_{b21}=Q_{c21}=0$. For this case the result resembles that for the model involving single AI level, discussed in \cite{RRZ06}. We observe only single transparency window, and its position and width do not depend on the value of $\Gamma_{21}$. We can only observe well-defined changes in the amplitude of variations of the real and imaginary parts of $\chi$, so the depth of the window becomes more distinct as $\Gamma_{21}$ increases. However, if we assume that $Q_{b21}$ and $Q_{c21}$ becomes different from zero (see Fig. 5) situation changes considerably again. Similarly as in Figs. 2 and 3, additional transparency window and region of anomalous dispersion appears again. From Fig. 5 we see that with increasing difference between the values of AI widths ($\Gamma_{21}$), the separation and widths of the both windows become more pronounced.
These facts justify the statements that the phenomena related to the autoionization processes strongly depend on the continuum shape and their characters. The strengths of the effects observed in the system can be changed considerably by varying the parameters describing the profile of the continuum, so we can have various possibilities of controlling these phenomena in practice.

\section{Conclusions}

In this paper we considered the $\Lambda$-like model involving two AI levels (for simplicity we assumed that they are of the same energy). This model is an extension of that with a single AI level, discussed by Raczy\'nski \textit{et.al} \cite{RRZ06}. For such a model we have derived the analytical formula describing the media susceptibility $\chi$. We have shown that due to the presence of the second AI level we can observe additional transparency window and extra region of the anomalous dispersion. We have shown that the properties (position, and width) of the window depend on the values of the parameters describing the interaction of these levels with the driving field. Moreover, the depth of the window can be manipulated by changes of the asymmetry parameters related to the transitions induced by the probe field and especially depends on the difference between the values of the autoionization widths. In addition, for the degenerate case if the parameters describing two AI levels are identical, our model behaves as that with one AI level characterized by some effective AI width and asymmetry parameter. If the parameters corresponding to the two AI levels start to differ each other, the additional EIT window appears despite the presence of degeneracy. This situation resembles that discussed in \cite{LTK87} concerning the long-time photoelectron spectra, when for various values of the AI level's parameters an additional zero appeared for the degenerate case. This is a result of the existence of two channels of autoionization and quantum interference between them. Such interference disappears if two AI levels are identical. We have shown that inclusion of additional AI state into the model can lead to the new and interesting effects that are absent for single level's models. 

The most important phenomenon discussed here is the appearance of the additional transparency windows in the system, where various channels of ionization (autoionization) exist. Such channels can interfere each other giving new EIT windows. These effects could seems to be similar to those observed in the systems involving only discrete levels but they are of completely different physical character. Indeed, for the model discussed here, we deal with a  structured continuum interacting with two discrete ground levels. In consequence, for the system considered here, we have  a new possibility to simultaneous slowing down of light pulses at various frequencies.

\bibliography{autojon}
\newpage
\label{Picture1}
\begin{center}
\includegraphics[width=6cm]{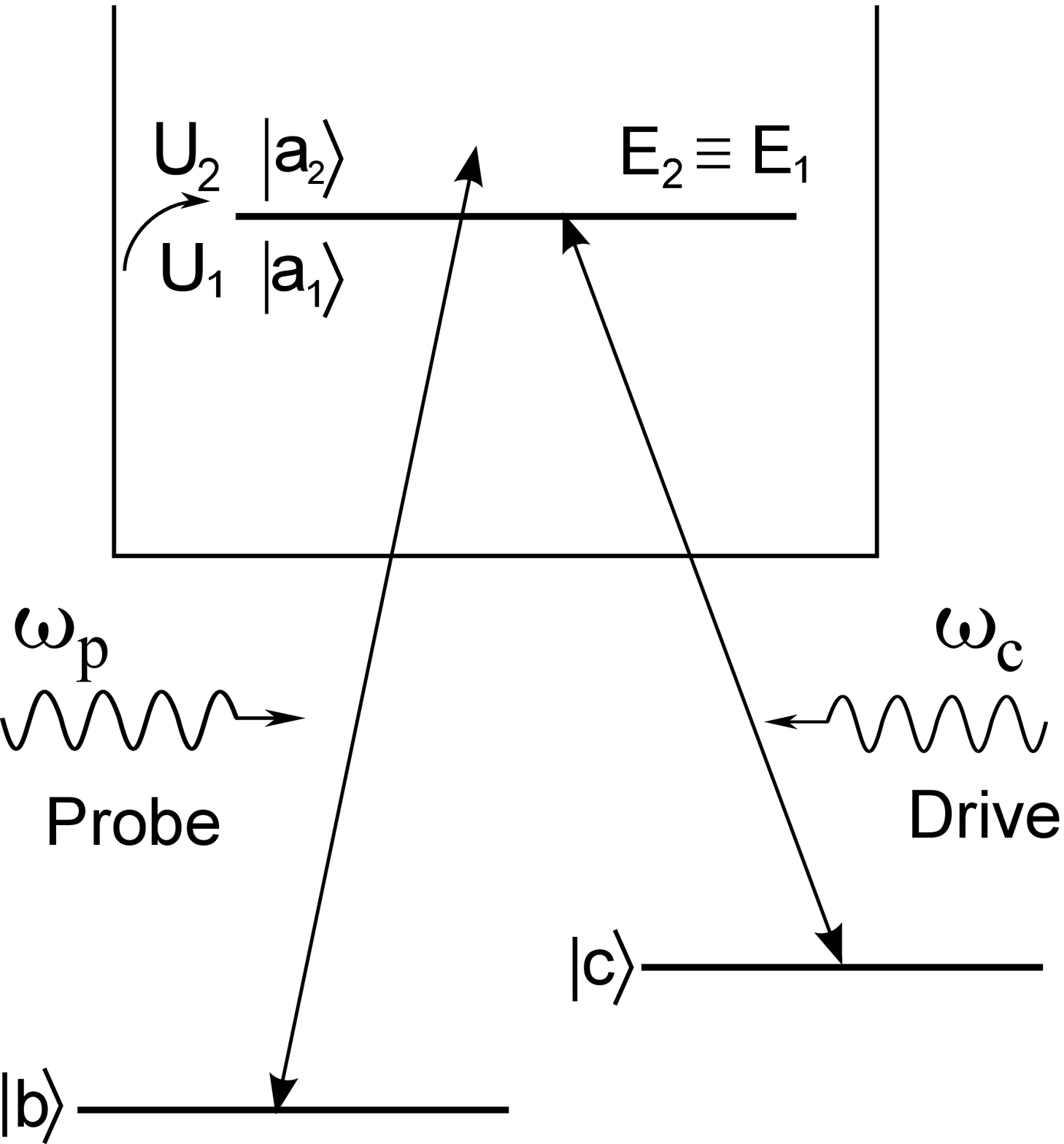}
\end{center}
\noindent\textbf{Fig.1}\quad{Scheme of the discussed model. Due to presence of the configurational interaction coupling ($U_1$ and $U_2$) between the two AI levels $|a_1\rangle$ and $|a_2\rangle$, and the flat continuum $|E\rangle$ all these states can be replaced by the double Fano structured continuum $|E)$. This continuum ($|E)$) is coupled by the weak probe and strong control fields of the frequencies $\omega_p$ and $\omega_c$, respectively.} 
\newpage
\label{Picture2}
\begin{center}
\includegraphics[width=14cm]{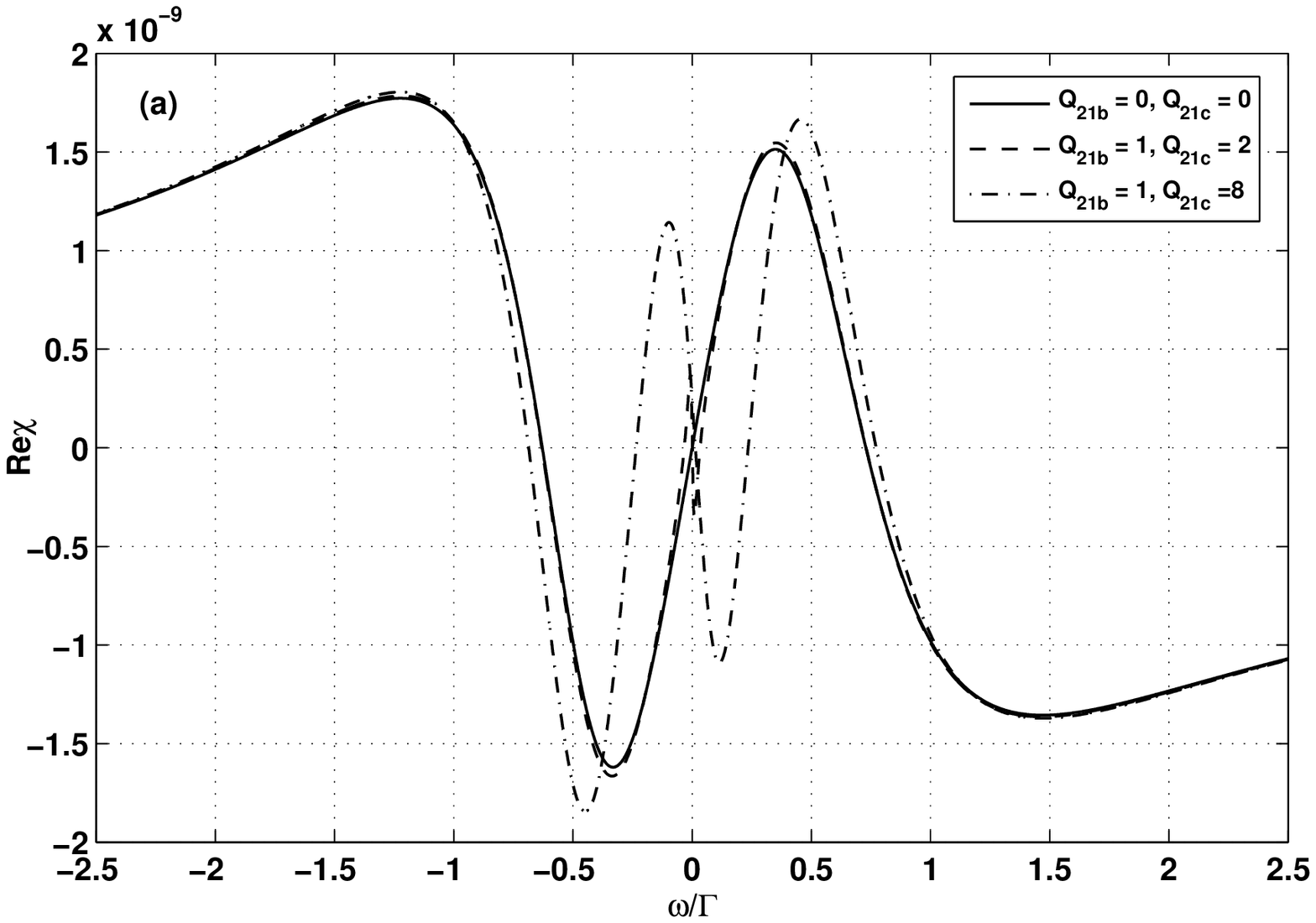}\\
\includegraphics[width=13cm]{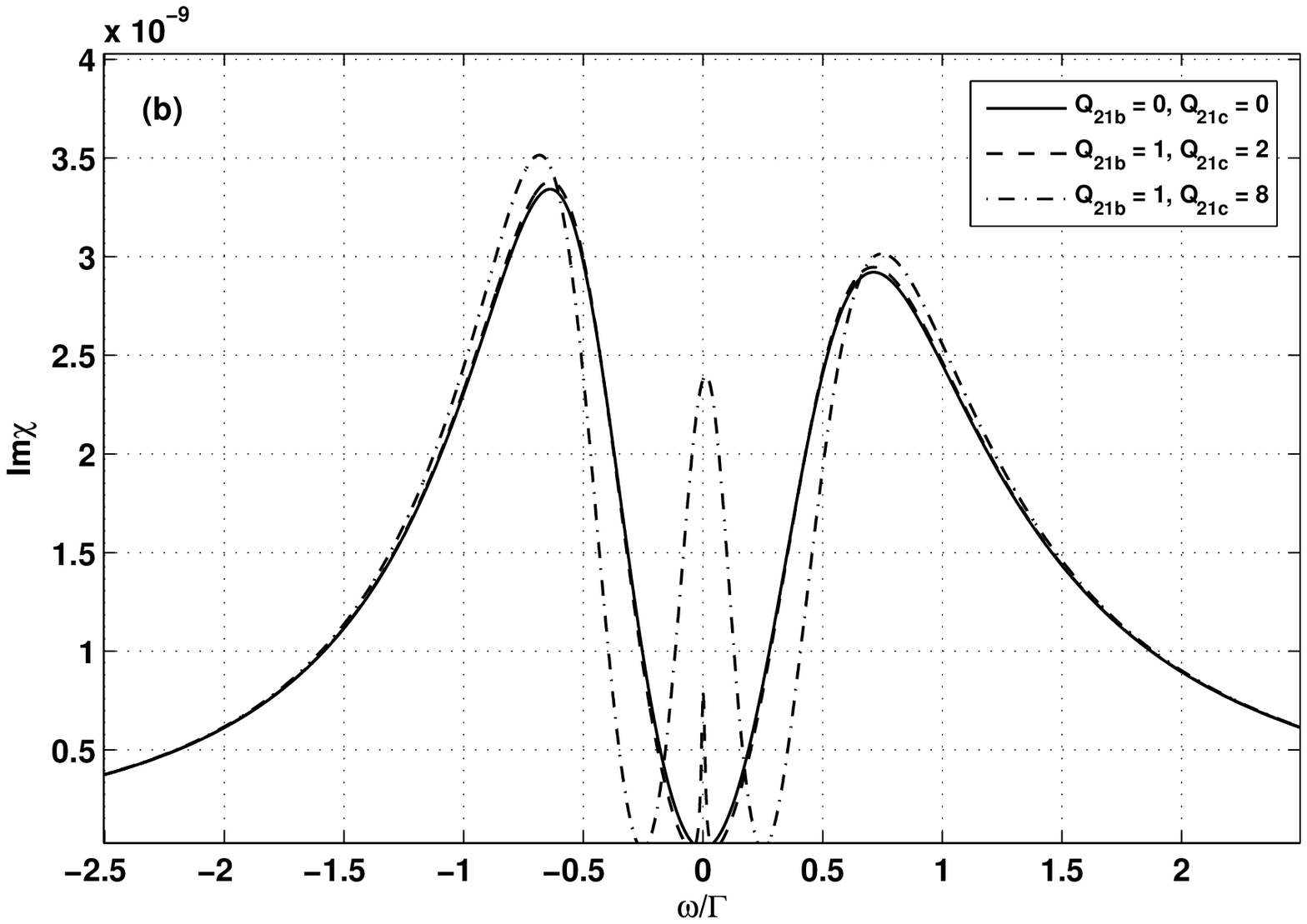}
\end{center}
\noindent \textbf{Fig.2}\quad{The real (a) and imaginary (b) parts of the susceptibility $\chi$ as a function of the detuning $\omega$ (in units of $\Gamma$). We assume that $\epsilon_{2}=4\times10^{-7}$ a.u., $\Gamma_{21}=0$ and $Q_b=Q_c=20$. Solid lines -- $Q_{b21}=Q_{c21}=0$, dashed lines -- $Q_{b21}=1,Q_{c21}=2$, -- dashed dotted lines $Q_{b21}=1$, $Q_{c21}=8$.}
\newpage
\label{Picture3}
\begin{center}
\includegraphics[width=12cm]{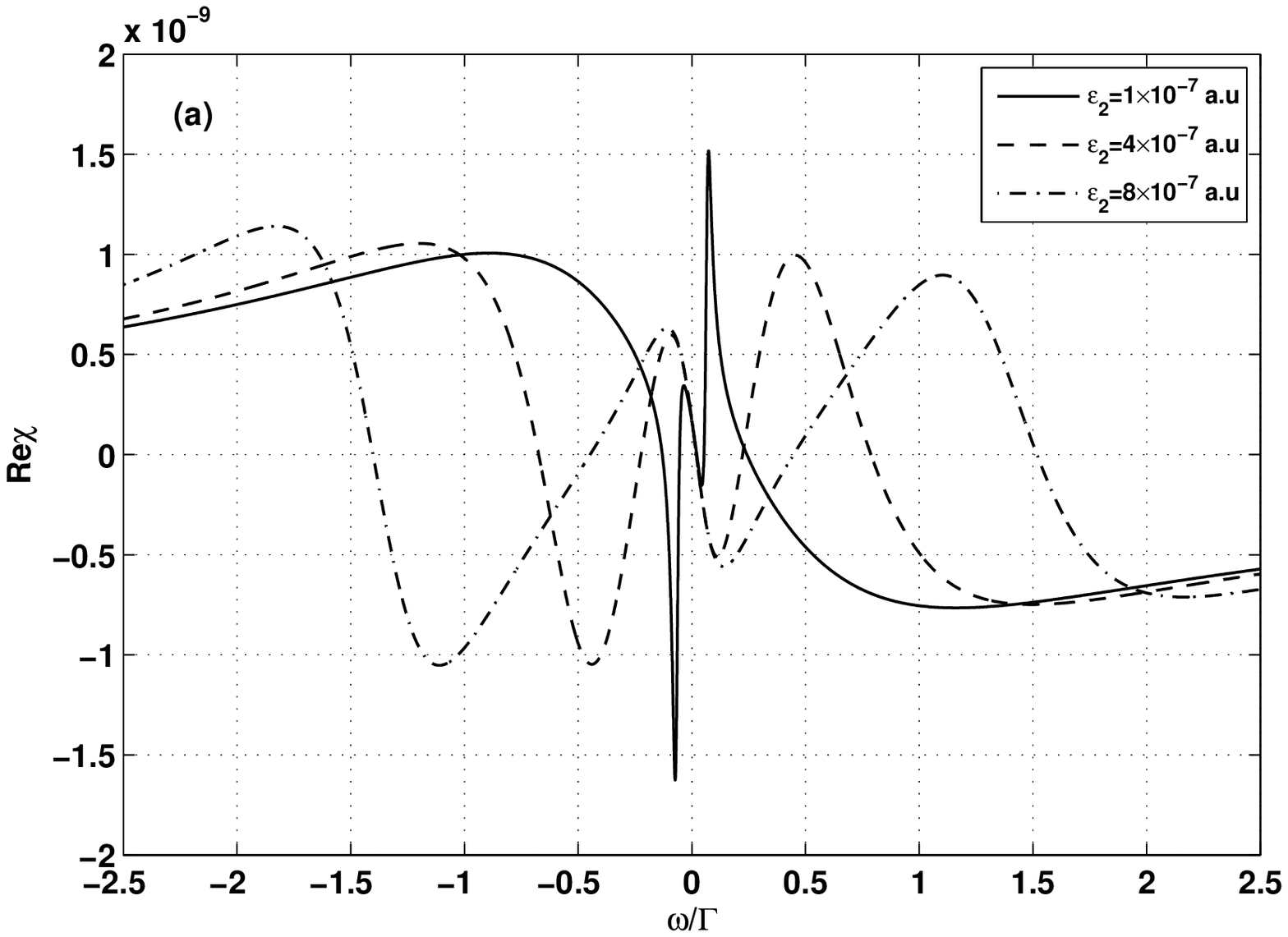}\\
\includegraphics[width=12cm]{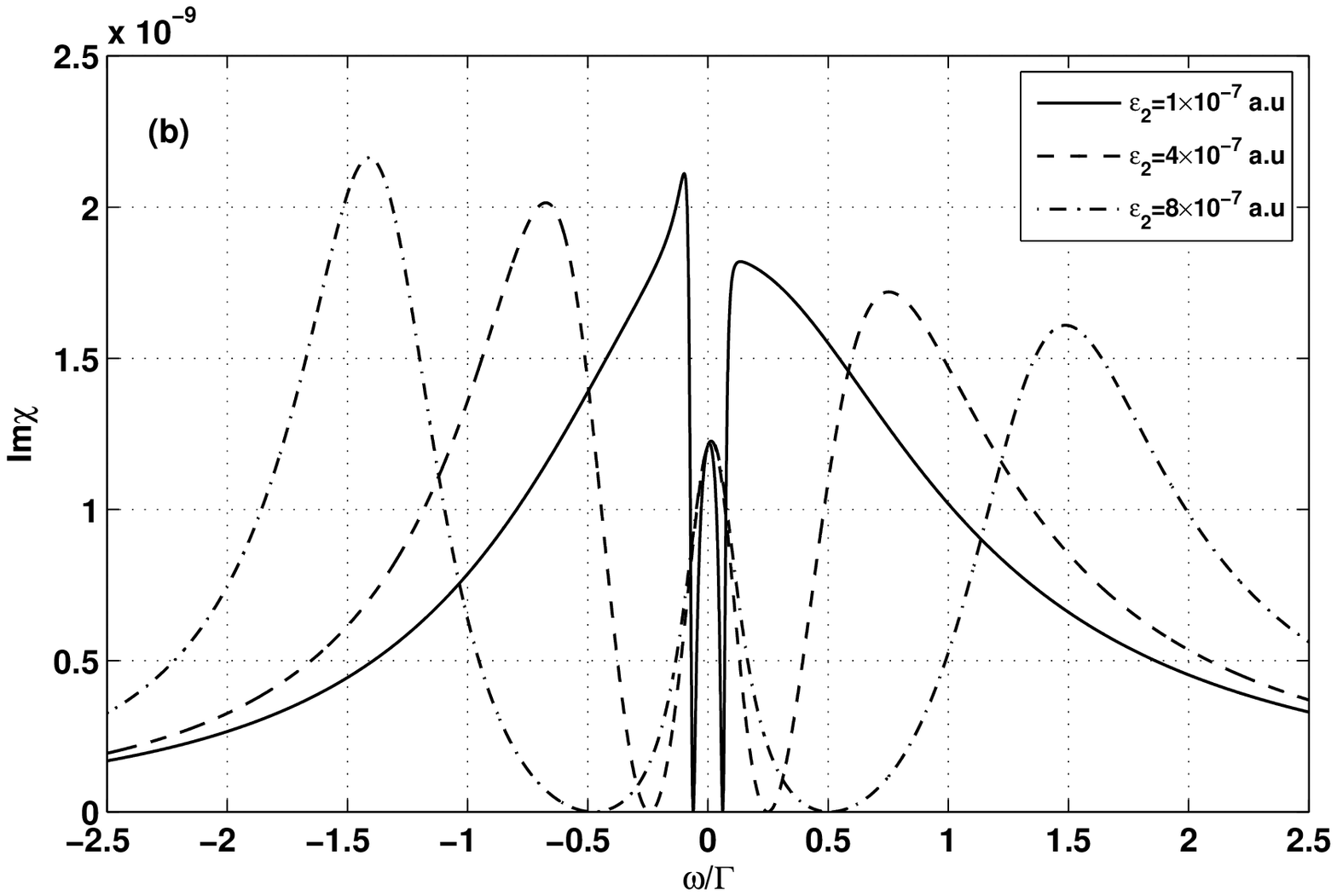}
\end{center}
\noindent\textbf{Fig.3. }{The same as in Fig.2 but for  $Q_{b21}=1, Q_{c21}=8$ and various values of $\epsilon_{2}$.  The remaining parameters are the same as in Fig.2.} 

\newpage
\label{Picture4}
\begin{center}
\includegraphics[width=12cm]{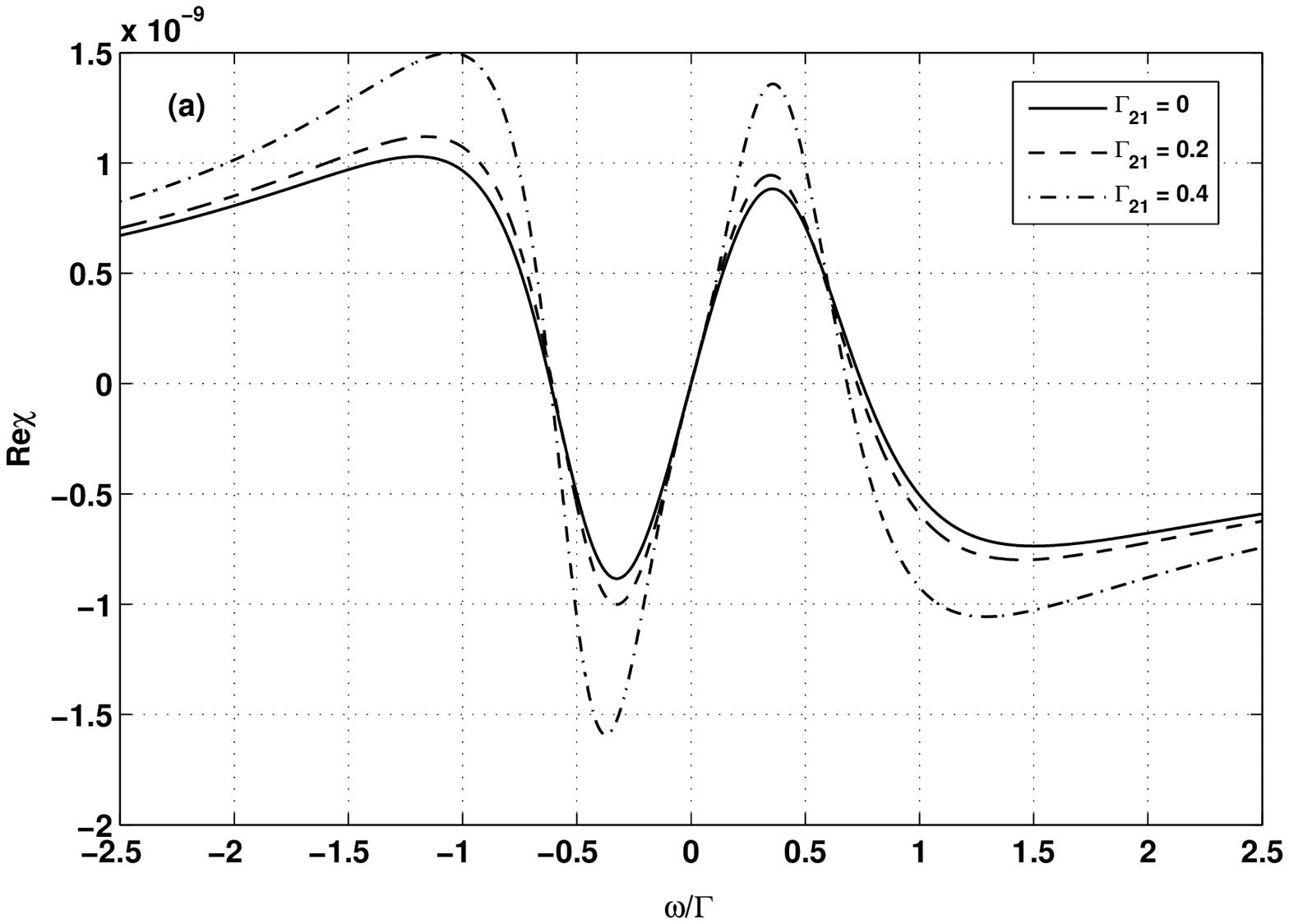}\\
\includegraphics[width=12cm]{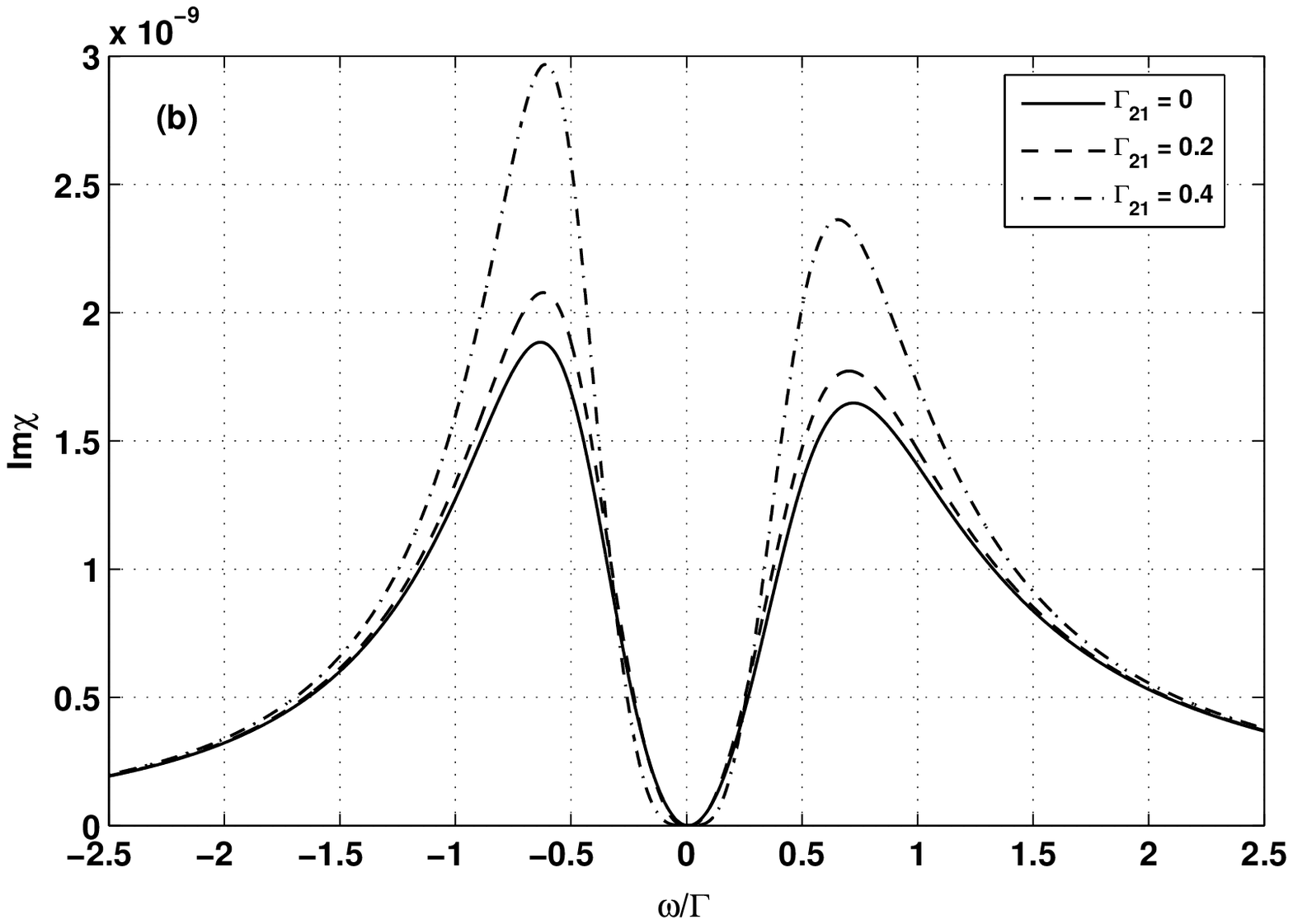}
\end{center}
\noindent\textbf{Fig.4. }{The real (a) and imaginary (b) parts of the susceptibility $\chi$ as a function of the detuning $\omega$ (in units of $\Gamma$)for identical AI levels ($Q_{b21}=Q_{c21}=0$) and various values of $\Gamma_{21}$.  We assume that $\epsilon_{2}=4\times10^{-7}$ a.u, and $Q_b=15, Q_c=20$.} 

\newpage
\label{Picture5}
\begin{center}
\includegraphics[width=12cm]{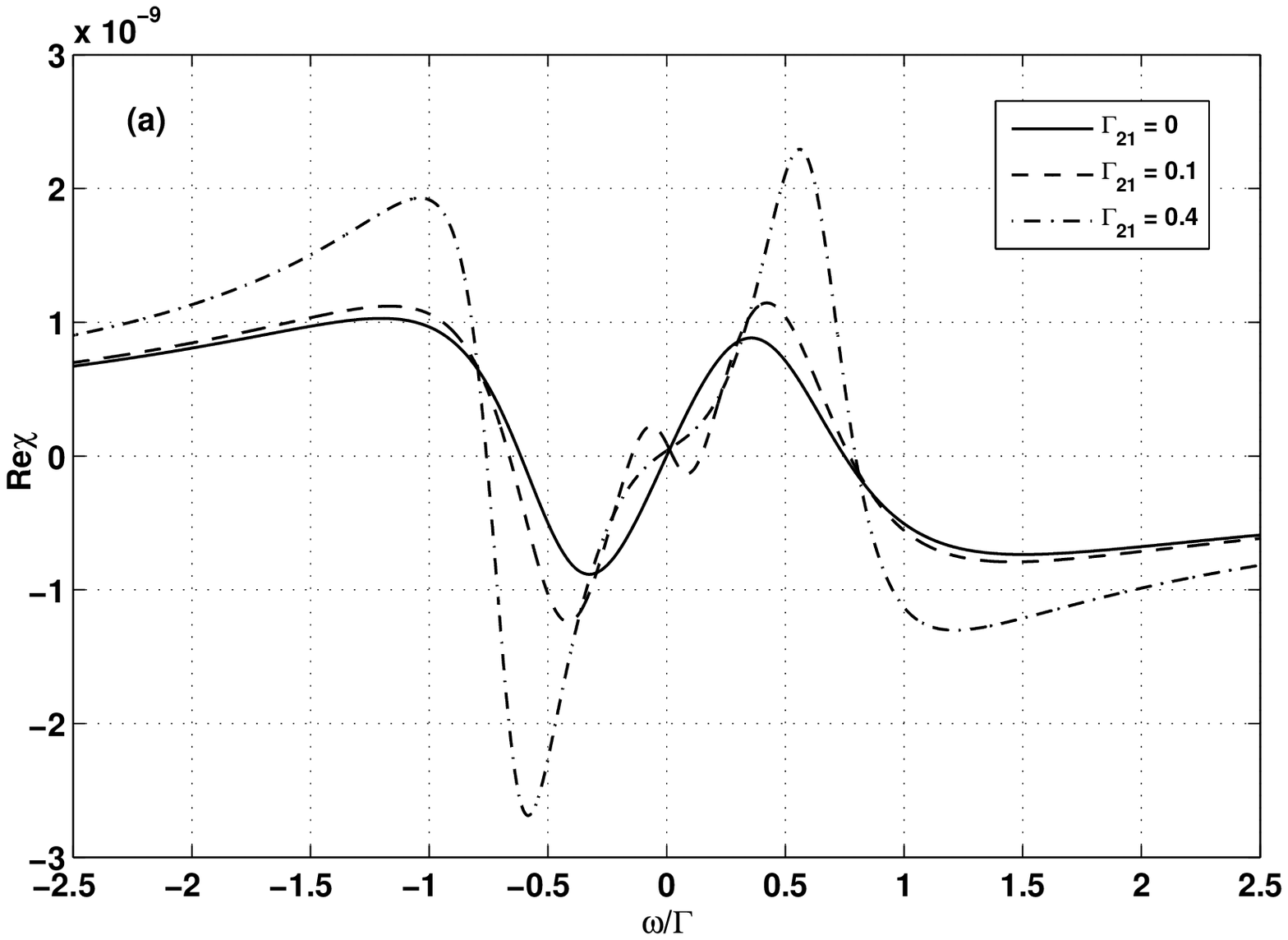}\\
\includegraphics[width=12cm]{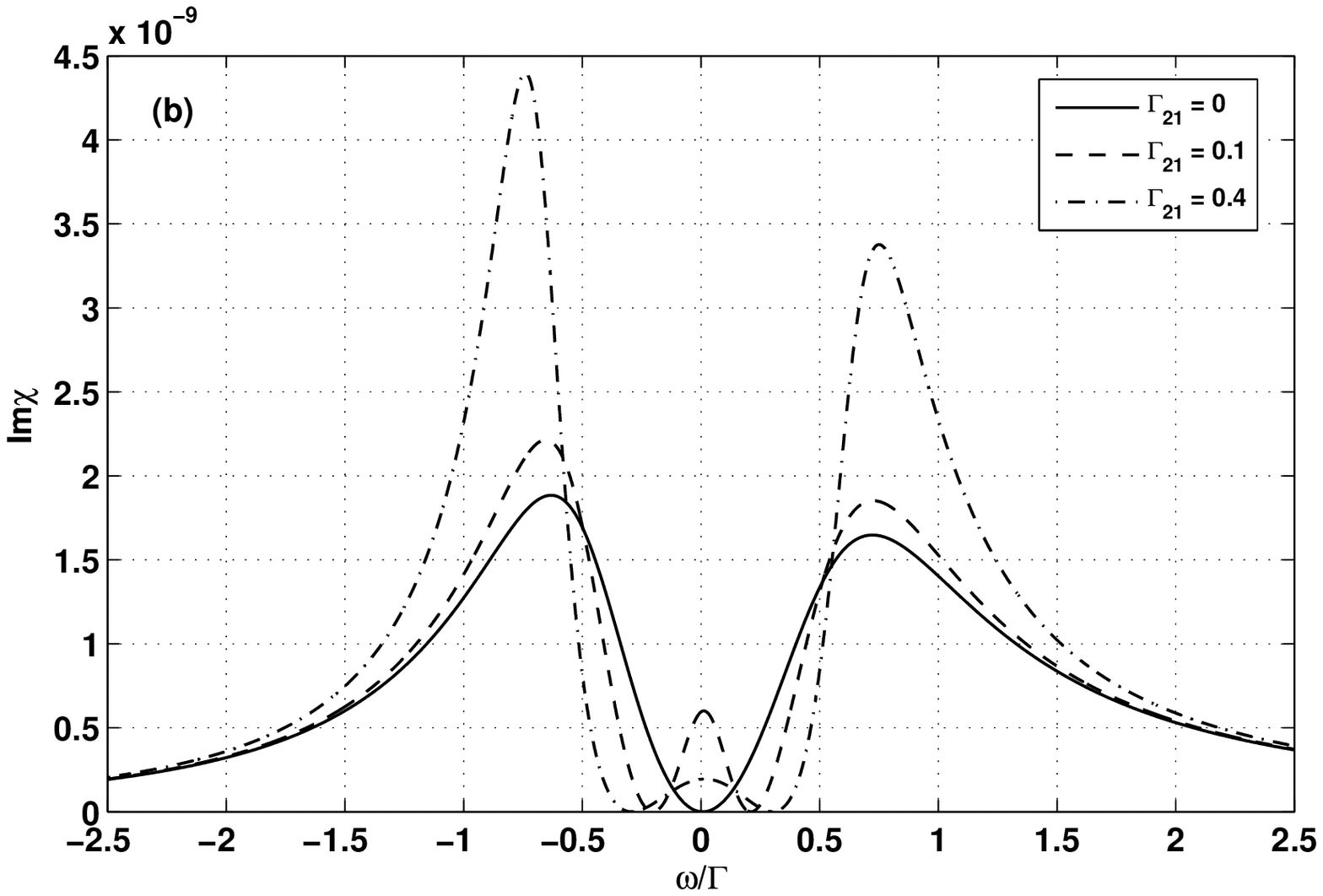}
\end{center}
\noindent\textbf{Fig.5. }{The real (a) and imaginary (b) parts of the susceptibility $\chi$  for various values of $\Gamma_{21}$, $Q_{b21}$ and $Q_{c21}$.  Solid lines -- $Q_{b21}= Q_{c21}=0 , \Gamma_{21}=0$, dashed lines -- $Q_{b21}=1$, $Q_{c21}=6, \Gamma_{21}=0.1$, -- dashed dotted lines $Q_{b21}=1$, $Q_{c21}=6$, $\Gamma_{21}=0.4$. The remaining parameters are the same as in Fig.4.}

\newpage
%\addtolength{\baselineskip}{-8truept}

\end{document}